\documentclass[pre,showpacs,a4paper,superscriptaddress,preprint,floatfix]{revtex4-2}
\usepackage[latin1]{inputenc}
\usepackage{amsmath}
\usepackage{amsfonts}
\usepackage{amssymb}
\usepackage{graphicx}
\usepackage{enumitem}
\usepackage[left=2cm,right=2cm,top=2cm,bottom=2cm]{geometry}
\usepackage{multirow}
\usepackage[normalem]{ulem}
\usepackage{rotating}
\useunder{\uline}{\ul}{}

\newcommand{\rme}{\mathrm{e}}
\newcommand{\rmc}{\mathrm{c}}
\newcommand{\rmi}{\mathrm{i}}
\newcommand{\rms}{\mathrm{s}}
\newcommand{\rmd}{\mathrm{d}}

\newcommand{\rmg}{\mathrm{g}}
\newcommand{\rmpl}{\mathrm{p}}
\newcommand{\rmrf}{\mathrm{rf}}
\newcommand{\rmdc}{\mathrm{dc}}
\newcommand{\rmde}{\mathrm{D_e}}

\begin{document}

\title{Stability of two-dimensional complex plasma monolayers in asymmetric capacitively-coupled radio-frequency discharges}

\author{L. Cou\"edel}
\affiliation{Physics and Engineering Physics Department, University of Sakatchewan, Saskatoon, SK S7N 5E2, Canada}
\affiliation{CNRS, Aix-Marseille Universit\'e, Laboratoire PIIM UMR 7345, 13397 Marseille cedex 20, France}
\email{lenaic.couedel@usask.ca, lenaic.couedel@univ-amu.fr}


\author{V. Nosenko}
\affiliation{Institut f\"ur Materialphysik im Weltraum, Deutsches Zentrum f\"ur Luft- und Raumfahrt (DLR), D-82234 We{\ss}ling, Germany}

\begin{abstract}
In this article, the stability of a complex plasma monolayer levitating 
in the sheath of the powered electrode of 
an asymmetric capacitively coupled radio-frequency argon discharge is studied. 
Compared to earlier studies, a better integration of the experimental results and theory 
is achieved by operating with actual experimental control parameters such as the 
gas pressure and the discharge power.
It is shown that for a given microparticle monolayer at a fixed discharge power there exist 
two threshold pressures: (i) above a specific pressure $p_\mathrm{cryst}$, 
the monolayer always crystallises; (ii) below a specific 
pressure $p_\mathrm{MCI}$, the crystalline monolayer undergoes the 
mode-coupling instability and the two-dimensional complex plasma 
crystal melts. In-between $p_\mathrm{MCI}$ and 
$p_\mathrm{cryst}$, the microparticle monolayer can be either in the fluid phase or the crystal 
phase: when increasing the pressure from below 
$p_\mathrm{MCI}$, the monolayer remains in the fluid phase until it 
reaches $p_\mathrm{cryst}$ at which it recrystallises;  when 
decreasing the pressure from above $p_\mathrm{cryst}$, the monolayer remains in the crystalline phase  
until it reaches $p_\mathrm{MCI}$ at which the mode-coupling 
instability is triggered and the crystal melts.
A simple self-consistent sheath model is used to calculate the rf sheath profile, the 
microparticle charges and the 
microparticle resonance frequency as a function of power and 
background argon pressure. Combined with calculation of the 
lattice modes the main trends of $p_{\rm MCI}$ as  a function of power and background argon pressure are 
recovered. The threshold of the mode-coupling instability in the 
crystalline phase is dominated by the crossing of the 
longitudinal in-plane lattice mode and the out-of plane lattice mode induced by the change of the 
sheath profile. Ion wakes are shown to have a significant effect too.
\end{abstract}

\maketitle

\section{Introduction}

Two-dimensional (2D) complex plasma crystals are composed 
of negatively charged monosized spherical 
microparticles levitating in the sheath above a confining 
electrode \cite{Morfill1996}. Complex plasma crystals are generally studied in asymmetric 
capacitively coupled-radio frequency (cc-rf) argon discharges 
in which the injected monolayer of microparticles 
levitates in the sheath above the powered electrode 
and crystallises under specific discharge conditions 
\cite{Melzer2000,Nunomura2000,Schweigert2002,Nosenko2003,Nunomura2000,
Nunomura2002,Qiao2003,Couedel2019,Couedel2010}.
Since microparticles can easily be imaged thanks to 
laser light scattering and the use of fast cameras, 
microparticle trajectories can be recovered and one can obtain 
information about the crystal at the kinetic "particle" level. Thus,
complex plasma crystals often serve as model systems to study 
generic phenomena such as wave propagation \cite{Nosenko2003,Samsonov2005}, 
shocks and phase transitions \cite{Thomas1996}.

However, due to the very nature of the complex plasma monolayer, complex plasma specific 
phenomena can occur. Because of the ion flow coming 
from the bulk plasma and directed towards the electrode, ion wakes are formed 
downstream of each microparticle \cite{Ishihara1997,Vladimirov2003,Matthews2020}. 
The ion wakes exert attractive force on the neighbouring particles 
making the microparticle-microparticle interactions non-reciprocal.
Under specific conditions, these non-reciprocal interactions can trigger the mode-coupling 
instability (MCI) \cite{Ivlev2001,Ivlev2003a,Zhdanov2009,Couedel2010,Couedel2011,Ivlev2014} 
in which energy from the flowing ions is transferred to the microparticles and can 
heat up the microparticle suspension.  Indeed, in 2D complex plasma crystals, 
three wave modes can be sustained: longitudinal and transverse in-plane 
acoustic modes and, due to the finite strength of the 
vertical confinement, an out-of-plane optical mode (associated 
with vertical oscillations \cite{Vladimirov1997,Ivlev2001,Qiao2003,Zhdanov2009,Couedel2009a}. 
Due to wake attraction, when the longitudinal in-plane mode and 
the out-of-plane mode cross, an unstable  hybrid mode is formed (the typical fingerprints being
a hot spot in reciprocal space at the edge of the first Brillouin zone, angular dependence and 
mixed polarisation \cite{Zhdanov2009,Couedel2011}), 
which can trigger MCI if the instability growth rate is larger than the 
damping rate due to friction with the neutral background \cite{Zhdanov2009}. 
MCI is observed in 
both crystalline \cite{Zhdanov2009,Couedel2010,Couedel2011} 
and fluid monolayers \cite{Ivlev2014,Yurchenko2017}. In the 
first case, MCI can result in rapid melting of the monolayer if its growth rate is high enough. 
In the latter case, MCI can prevent crystallisation 
of the monolayer by pumping more kinetic energy into the microparticles than the energy dissipated through 
friction with the neutral gas background \cite{Williams2012,Roecker2014,Yurchenko2017}.  
Since  for given
 discharge parameters (rf power, neutral gas pressure, 
 geometry) and fixed monolayer parameters (number density 
 and size of the microparticles) the growth rate in a fluid monolayer 
is higher than in a crystalline monolayer\cite{Ivlev2014}, it allows 
external triggering of the MCI \cite{Couedel2018}  and 
prevents recrystallisation of a melted monolayer without strongly altering the 
discharge parameters \cite{Couedel2018,Yurchenko2017}.

It was shown experimentally that MCI is triggered at specific pressure and rf power 
thresholds \cite{Couedel2010,Couedel2011}. 
MCI is very sensitive to the magnitude of  the effective wake dipole moment since 
the shape of the dispersion relations critically depends on it \cite{Rocker2014}.
MCI threshold and  growth rate 
also strongly depend on the depth of the mode crossing \cite{Rocker2014}, 
which is a function the ion-wake parameters and vertical confinement frequency. 
Both the ion wake and the vertical confinement frequency depend on the parameters 
of the rf sheath in which the microparticles are levitating such as the 
strength of the confining sheath electric field and the local ion and electron densities. 
Therefore, in order to understand the dynamics of the microparticle 
monolayer, it is very important to be able to relate the discharge parameters (background 
argon pressure, rf power) to the sheath parameters (electric field profile, ion 
flux profile, ion and electron density profiles) . Modeling the rf sheath in a 
cc-rf discharge is however not trivial and accurate 
calculation of rf sheath profile has been a subject of active research for the 
past three decades, see for example Refs.~\cite{Lieberman1988,Lieberman1989,Godyak1990a,Chandhok1998,
Gierling1998,Sobolewski2000,Dewan2002,Chabert2011,Czarnetzki2013,Chabert2021}.

In this article, the stability of  2D 
complex plasma monolayers levitating in the sheath above the powered 
electrode of a cc-rf discharge is  experimentally studied. 
In Sec.~\ref{ExpSetUp}, the experimental set up is described.
In Sec.~\ref{sec:Electrical}, the experimental characterisation of 
the cc-rf discharge (electron density and 
powered electrode self bias) as a function of the 
background argon pressure and the forward rf power is presented. 
In Sec.~\ref{expMCI}, a study of the stability (crystal or fluid) of specific microparticle 
monolayers as a function of rf power is performed. It allowed us to 
measure the MCI pressure threshold and the crystallisation pressure threshold over 
a wide range of rf powers.
In Sec.~\ref{SheathModel}, a simple self-consistent sheath model is described and used to calculate 
the sheath profile and microparticle charge, and vertical resonance frequency as a function of 
discharge parameters. In Sec.~\ref{MCI}, the results of Sec.~\ref{SheathModel} are used to 
calculate the lattice modes of 2D complex plasma crystal with and without ion wakes. The influence 
of the interparticle distance and wake 
parameters are investigated. The calculated results are compared to experimental results. 
In Sec.~\ref{Conclusion}, the main results are summarised and future investigations are discussed.

\section{Experimental set up}\label{ExpSetUp}

\begin{figure}[htbp]
    \centering
    \includegraphics[width=3.0in]{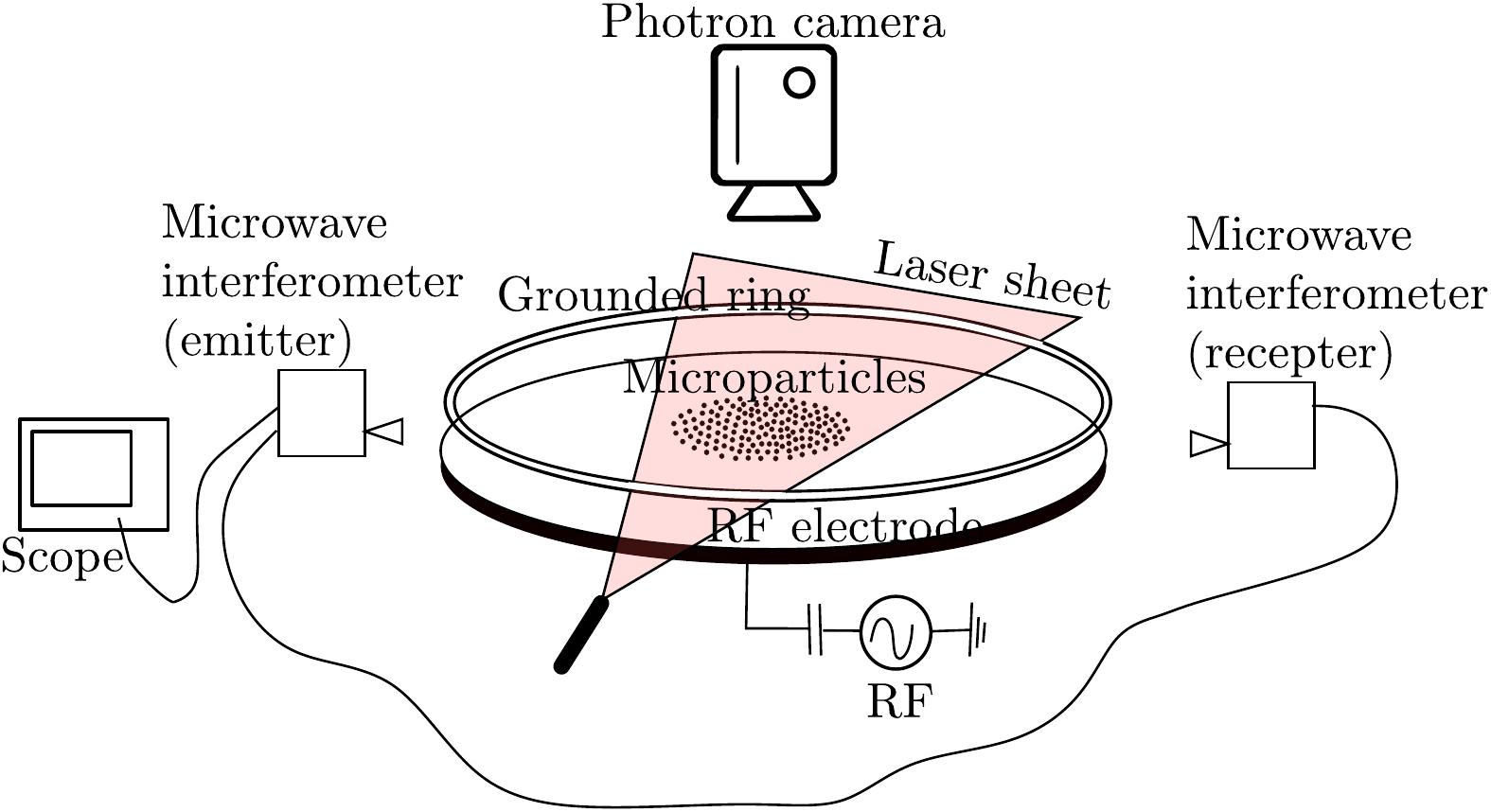}
    \caption{Schematic of the experimental set up. Note that the microwave interferometer 
    was used only for measurements in  pristine plasmas (i.e. no microparticles)}
    \label{fig:setup}
\end{figure}

The experiments were performed in a modified Gaseous Electronic Conference 
rf cell (GEC). It is an asymmetric cc-rf discharge at 13.56~MHz. A 
sketch of the setup is shown in Fig.~\ref{fig:setup}. 
Experiments were performed with an argon pressure, $p_\mathrm{Ar}$, between
0.5 and 2 Pa. The forward rf power, $P_\mathrm{W}$, was set between 5 and 25 W. 
Langmuir probe studies using the same setup have shown that the electron temperature is
$T_\rme \sim 2.5$~eV and that the electron density is $n_\rme \simeq 2 \cdot 10^9\ \mathrm{cm^{-3}}$ at
$p_\mathrm{Ar} = 0.66$~Pa and $P_{\rm W}= 20$~W~\cite{Nosenko2009}. 
A Miwitron Ka-band microwave 
interferometer MWI 2650 working at 26.5~
GHz was used to measure the electron density, $n_{\rm e}$, in pristine plasma
condition (without injected microparticles) for different $p_{\rm Ar}$ and $P_{\rm W}$. 
The interferometer had a resolution 
of ${\sim 10^8\ \mathrm{cm^{-3}}}$.

Melamine-formaldehyde (MF) spherical microparticles with 
a diameter of $9.19 \pm 0.09\ \mathrm{\mu m}$ were
injected in the plasma. They levitated in the sheath above the lower rf electrode 
where the electric force 
balanced their weight and formed a highly ordered horizontal monolayer.
The microparticle monolayer was illuminated by a horizontal
laser sheet and the microparticles were imaged 
through the top chamber window by using a 
4 Megapixel Photron FASTCAM Mini WX100 camera at a speed of 250 frames per second.
Particle tracking allowed us to recover the particle horizontal coordinates, 
$x$ and $y$ with subpixel resolution in each
frame, and the velocities, $v_x$ and $v_y$, were then calculated \cite{Couedel2019}. 
An additional side-view camera (Basler Ace ACA640-
100GM) was used to check that no particles were levitating above or under the main monolayer. 
More details can be found in previous publications 
\cite{Nosenko2006,Nosenko2007,Nosenko2009,Couedel2010}

\section{Characterisation of the cc-rf discharge}\label{sec:Electrical}

\begin{figure}[htbp]
    \centering
    \includegraphics[width=0.95\columnwidth,height=73.4mm,keepaspectratio]{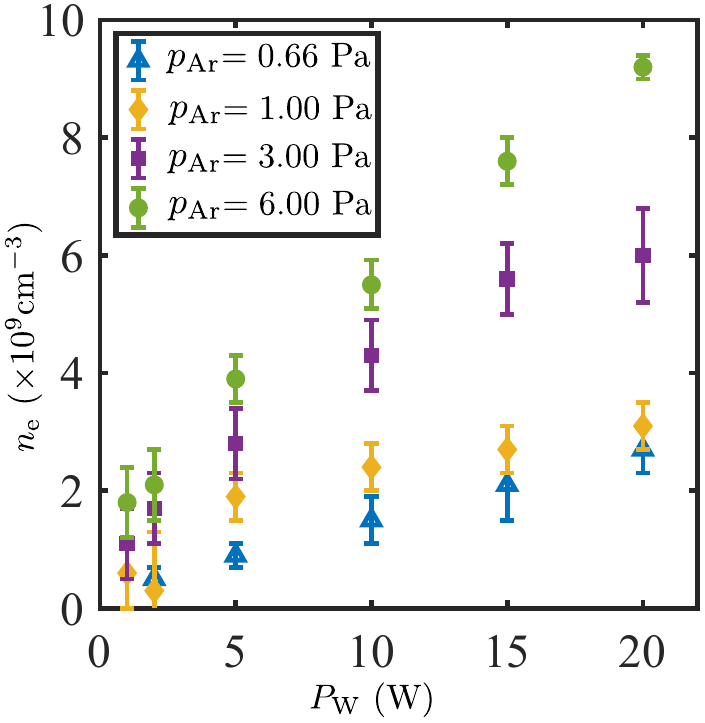}
    \caption{(Colour online) Electron density, $n_\rme$, as a function 
    of RF power, $P_W$, for different pressures}
    \label{fig:ne_exp}
\end{figure}

Electron density measurements have been taken with the microwave 
interferometer for different pressure between 
0.66~Pa and 6.0~Pa and for rf power between 1~W and 20~W. The results are are shown in
Fig.~\ref{fig:ne_exp}. The density reported at 
$p_{\rm Ar}=0.66$~Pa and $P_{\rm W}=20$~W is 
$n_\rme \sim 2.5 \cdot 10^9 \ {\rm cm^{-3}}$ which is in agreement with 
previous Langmuir probe measurements \cite{Nosenko2009}. As expected, $n_\rme$ increases with  
forward RF power and  pressure \cite{Lieberman1994,Chabert2011}. The measured values of bulk plasma density 
are used as input parameters for sheath profile calculations in Sec.~\ref{CalculatedProfiles}.

\begin{figure}[htbp]
    \centering
    \includegraphics[width=0.99\columnwidth,height=48.7mm,keepaspectratio]{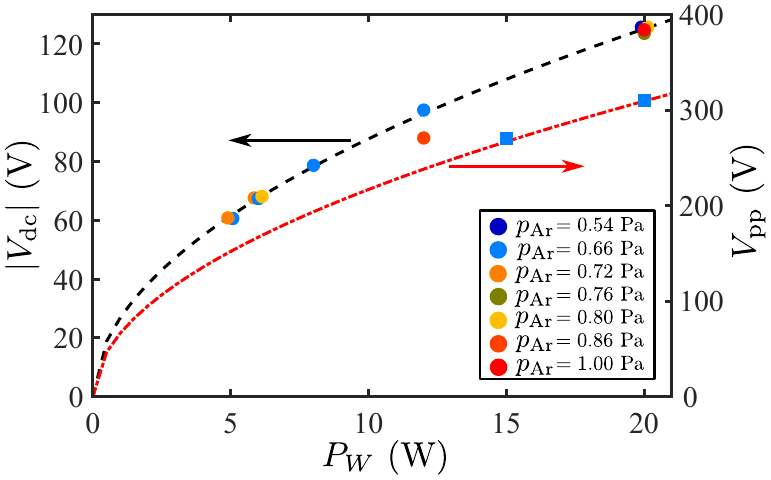}
    \caption{(Colour online) ($\circ$) Self-bias voltage $|V_{\rm dc}|$ as a function 
    of RF power, $P_W$ for different working pressures 
    (colour-coded). ($\square$) Peak-to-peak RF voltage,  
    $V_{pp}$, as a function or RF power, $P_W$ at $p_{\rm Ar}=0.66$~Pa. The dashed black line is 
    the fit $|V_{\rm dc}|= 27 \cdot P_{\rm W}^{0.51}$. The dotted red line is $V_{pp}$ calculated from 
    Eq.~(\ref{eq:vdc}).}
    \label{fig:Vdc}
\end{figure}

The absolute value of self-bias voltage on the powered electrode, $|V_{\rm dc}|$, 
has also been measured for a few relevant experimental discharge conditions.
The results are shown in Fig.~\ref{fig:Vdc} (dots). As can be seen, $|V_{\rm dc}|$ is not very sensitive 
to pressure. It has approximately a square-root dependence on the forward rf power.

It is known that the self-bias voltage is related to 
the ratio of the areas of the powered surface to  the grounded surface interacting 
with the plasma $A_\rmrf / A_\rmg$ \cite{Raizer1995,Lieberman1994}. 
The ratio of the self bias voltage $V_\rmdc$ to the amplitude of 
the rf voltage $V_0=V_{\rm pp}/2$ is in first approximation given by~\cite{Song1990,Land2009}:
 \begin{equation}\label{eq:vdc}
    \frac{V_\rmdc}{V_0}=\sin \Bigg(\frac{\pi}{2} 
    \Big(\frac{A_\rmg - A_\rmrf}{ A_\rmg + A_\rmrf} \Big) \Bigg),
\end{equation}
where $V_{\rm pp}$ is the peak-to-peak rf voltage.
It was experimentally measured that in a modified GEC cell similar 
to the one that was used for the current set of experiments,  
$A_\rmrf / A_\rmg \sim 0.25$ \cite{Land2009}. Using the fit of 
$V_{\rm dc}$ as a function of $P_W$ and Eq.~(\ref{eq:vdc}), 
$V_{\rm pp}$ was calculated and the result is shown by the 
dotted red line in Fig.~\ref{fig:Vdc}. The results match 
reasonably well with two measured values of $V_{pp}$ at 
15~W and 20~W for $p=0.66$~Pa (blue squares). The calculated values of $V_{\rm dc}$ and  $V_{\rm pp}$ 
are used as input parameters to calculate sheath profiles in Sec.~\ref{CalculatedProfiles}.

\section{Stability of complex plasma monolayers}\label{expMCI}

A large monolayer suspension was created by injecting  monosized 
melamine formaldehyde microparticles 
(diameter $9.19 \pm 0.09\ \mathrm{\mu m}$). The monolayer was kept for the 2 sets of 
experiments (referred to as Experiment~I and Experiment~II) separated by 
$\sim$1 hour. According to Ref.\cite{Kohlmann2019}, the 
MF microparticles are etched at a rate of $\sim 1.25\ {\rm nm/min}$ in an argon discharge 
meaning that the microparticle diameter for the second set of 
experiments was $\sim 9\ {\rm \mu m}$.

The crystallisation and melting pressures of the monolayer 
were explored for different rf powers.
At a given rf power, starting from a fluid monolayer, the argon pressure was 
gradually increased until the monolayer was fully 
crystallised and the crystallisation pressure was recorded. Note that during the 
pressure increase period, the monolayer remained in the \emph{fluid phase} \footnote{
Except when within a few $10^{-2}$~Pa of the crystallisation pressure where 
the outer part of the monolayer has started crystallising while the centre remained melted}. 
Then the pressure was very slowly 
decreased until MCI was triggered in the crystal resulting in  the rapid melting 
of the monolayer in a pattern 
similar to the results reported 
in Refs.~\cite{Couedel2014a,Laut2015,Yurchenko2017}. Note that during the pressure decrease 
period, the  monolayer remained in the \emph{crystalline phase} until MCI was triggered. 
The rf power was then decreased and the procedure was repeated.

\begin{figure}[htbp]
    \centering
    \includegraphics[width=60mm]{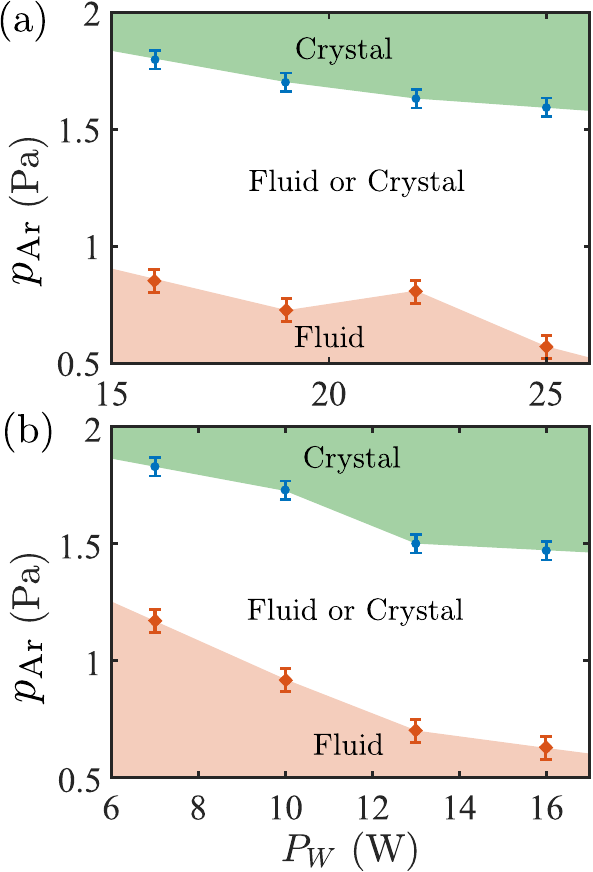}
    \caption{(Colour online) Crystallisation pressure, $p_{\rm cryst}$,  (blue dots) and 
    MCI threshold pressure, $p_{\rm MCI}$, (red diamonds) for 
    different rf powers, $P_{\rm W}$: (a) Experiment I; (b) 
    Experiment II. Above the crystallisation pressure
    the monolayer is always in the crystalline state (green areas). Below the MCI threshold pressure, the 
    monolayer is always in the fluid state. In-between, the monolayer remains in the crystalline phase 
    when decreasing the pressure from  above the crystallisation thershold and 
    remains fluid when increasing the pressure 
    from bellow the MCI threshold.}
    \label{fig:exp_stability}
\end{figure}

The results for the two sets of experiments are shown in Fig.~\ref{fig:exp_stability}. As can be seen, 
in both cases, the higher the power the lower the crystallisation and melting pressures. Moreover, the 
gap between the two pressures, in which the monolayer remains in the crystalline phase when 
decreasing the pressure from above the crystallisation threshold and  
remains fluid when increasing the pressure 
from bellow the MCI melting threshold, increases when increasing the RF power.

\begin{figure}[htbp]
    \centering
    \includegraphics[scale=0.75]{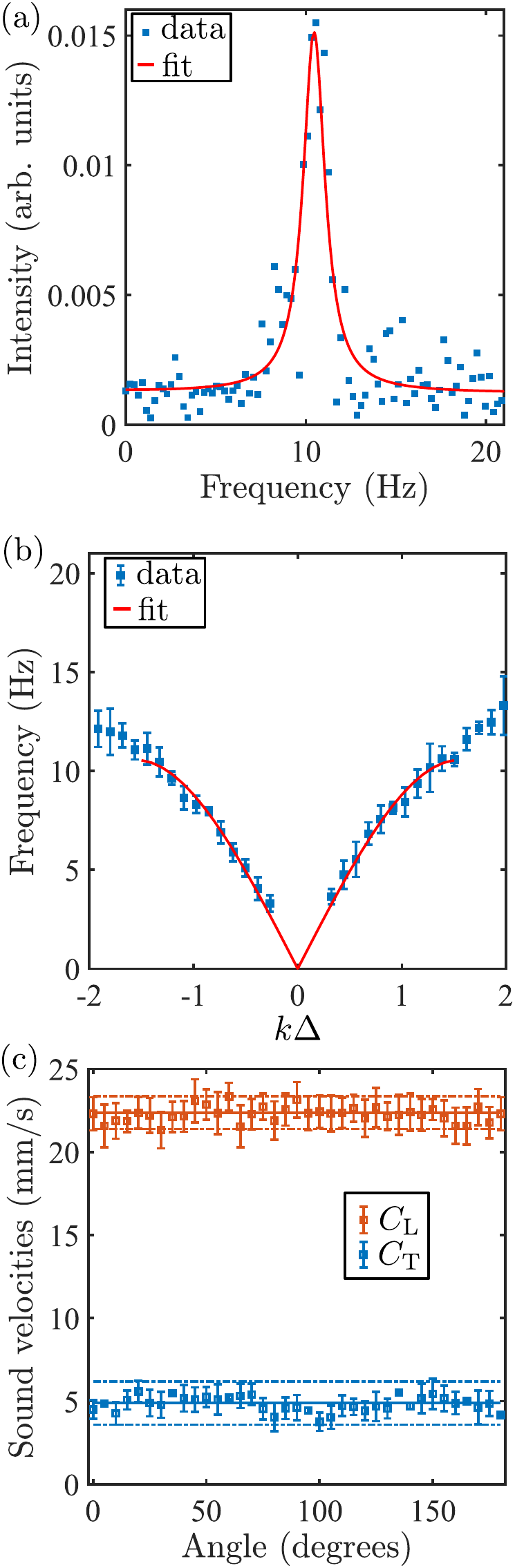}
    \caption{(a) Fit of the longitudinal current fluctuation spectrum for $k\Delta=1.3$. Only the 
    ``positive'' frequency side of the Fourier transform is shown. (b) Fit of the reconstructed 
    longitudinal dispersion relation for wave propagating at an angle of $0^{\circ}$ with 
    respect to the camera field of view. (c) Longitudinal sound speed, $C_{\rm L}$ and 
    transverse sound speed, $C_{\rm T}$, measured for different wave propagation angles   
    with respect to the camera field of view. All plots correspond to measurements 
    at $P_W=16$~W and $p=1.80$~Pa.
    }
    \label{fig:procedure_spectra}
\end{figure}

During the decreasing pressure phase, the crystalline monolayer was imaged with the 
high-speed Photron camera in order to track the particle trajectories and recover the 
longitudinal and transverse current fluctuation spectra following the procedure described in 
Refs.~\cite{Hansen2006,Liu2009a,Couedel2014a}. To extract the longitudinal and 
transverse dispersion relations $\omega_{\rm L,T}(k)= 2 \pi f_{\rm L,T}(k)$ 
where $f_{\rm L,T}(k)$ is the frequency of the lattice modes with 
wave number $k$, the
current fluctuation spectra $J_{\rm L,T} (k, f)$ were 
fitted to a double-Lorentzian form \cite{Kryuchkov2019,Khrapak2020}:
\begin{align}
    J_{\rm L,T} (k, \omega) \propto  & 
     \frac{A(k)}{(\omega -\omega_{\rm L,T})^2 + A(k)} \nonumber\\ + 
      & \frac{A(k)}{(\omega +\omega_{\rm L,T})^2 + A(k)},
\end{align}
with $A(k)$ a constant 
proportional to the damping rate of the mode. The indices $L$ and $T$ are for longitudinal 
and transverse, respectively. In Fig.~\ref{fig:procedure_spectra}(a), the fit of the 
current spectrum for a normalised wave number $k\Delta=1.3$, where $\Delta$ is the interparticle 
distance is shown and the entire experimental longitudinal dispersion relation is 
shown in Fig.~\ref{fig:procedure_spectra}(b) for wave propagating at an 
angle of $0^{\circ}$ with respect to the camera field of view. The interparticle distance $\Delta$ 
was obtained from the position of the first peak of the pair correlation 
function and error bars are the widths of the peaks at half maximum. 
Due to high noise level 
for $k\Delta<0.25$, the frequencies of the mode at these long wavelengths  were not measured. 
Then,  longitudinal and transverse dispersion relations were fitted for $k\Delta<1.5$ 
by the polynomial of the form
$\omega_{\rm L,T}= C_{\rm L,T} |k| + b_{\rm L,T} |k|^3$, where $C_{L,T}$ are 
the longitudinal and transverse sound speeds and 
$b_{\rm L,T}$ are constants taking into account 
the non-linearity of the dispersion relation for $|k\Delta| \gtrsim 1$. 

By numerically rotating the frame  with respect 
to the original camera field of view and repeating the procedure, wave propagation could be 
studied at different angles. $C_{L,T}$ were obtained 
every $5^{\circ}$ (see Fig.~\ref{fig:procedure_spectra}(c)) and the 
average values of the sound speeds were calculated for different crystal conditions.
Assuming screened Coulomb interactions and following the method of Ref.~\cite{Nunomura2002}, 
these values were then used to 
extract the microparticle charge $Z_{\rm d}$, the Debye 
(screening) length $\lambda_{\rmd}$ and the coupling 
parameter $\kappa=\Delta/\lambda_\rmd$. The results are 
summarised in Table~\ref{tab:crystal_parameters}.
\begin{table*}[!ht]
\caption{Two-dimensional crystal parameters for different discharge conditions between the crystallisation pressure and 
the MCI threshold pressure. The microparticle charge $Z_{\rm d}$, the Debye 
(screening) length $\lambda_{\rmd}$ and the coupling 
parameter $\kappa=\Delta/\lambda_\rmd$ were calculated assuming screened 
Coulomb interactions \cite{Nunomura2002}.}
\label{tab:crystal_parameters}
\begin{tabular}{lcccccccccc}
\hline \hline
                                                     & $P_{\rm W}$ & $p_{\rm cryst}$  & $p_{\rm MCI}$     & $p_{\rm Ar}$   & $\Delta$               & $C_{\rm L}$      & $C_{\rm T}$     & $Z_d$            & $\kappa$      & $\lambda_{\rm D}$ \\
{\ul }                                               & (W)         & (Pa)             & (Pa)              & (Pa)  & (mm)              & (mm/s)           & (mm/s)          & (e)              & {\ul }        & ($\rm \mu m$)     \\ \hline
\multicolumn{1}{l|}{\multirow{13}{*}{\begin{turn}{90}Experiment I\end{turn}}}  & 25          & $1.595 \pm 0.05$ & $0.575 \pm 0.033$ & 1.583 & $0.361 \pm 0.020$ & $21.53 \pm 1.66$ & $4.94 \pm 3.24$ & $10500 \pm 2900$ & $1.3 \pm 0.5$ & $330 \pm 180$     \\
\multicolumn{1}{l|}{}                                &             &                  &                   & 1.167 & $0.372 \pm 0.014$ & $22.23 \pm 1.79$ & $4.82 \pm 2.88$ & $10100 \pm 2400$ & $1.2 \pm 0.4$ & $370 \pm 170$     \\
\multicolumn{1}{l|}{}                                &             &                  &                   & 0.750 & $0.391 \pm 0.027$ & $25.60 \pm 2.16$ & $6.51 \pm 2.91$ & $14600 \pm 3000$ & $1.6 \pm 0.4$ & $270 \pm 90$      \\
\multicolumn{1}{l|}{}                                & 22          & $1.633 \pm 0.05$ & $0.808 \pm 0.033$ & 1.633 & $0.352 \pm 0.021$ & $22.44 \pm 2.35$ & $5.31 \pm 3.25$ & $11100 \pm 3000$ & $1.4 \pm 0.5$ & $300 \pm 150$     \\
\multicolumn{1}{l|}{}                                &             &                  &                   & 1.167 & $0.362 \pm 0.013$ & $23.04 \pm 2.53$ & $6.23 \pm 3.90$ & $14300 \pm 5100$ & $1.8 \pm 0.8$ & $240 \pm 120$     \\
\multicolumn{1}{l|}{}                                &             &                  &                   & 0.833 & $0.376 \pm 0.015$ & $26.50 \pm 2.58$  & $6.29 \pm 5.37$ & $14100 \pm 6000$ & $1.5 \pm 0.8$ & $470 \pm 390$     \\
\multicolumn{1}{l|}{}                                & 19          & $1.703 \pm 0.05$ & $0.725 \pm 0.017$ & 1.703 & $0.348 \pm 0.015$ & $21.41 \pm 1.47$ & $4.89 \pm 2.44$ & $10000 \pm 2100$ & $1.3 \pm 0.4$ & $300 \pm 110$     \\
\multicolumn{1}{l|}{}                                &             &                  &                   & 1.167 & $0.363 \pm 0.016$ & $25.42 \pm 1.05$ & $5.63 \pm 2.76$ & $11700 \pm 2300$ & $1.2 \pm 0.4$ & $330 \pm 120$     \\
\multicolumn{1}{l|}{}                                &             &                  &                   & 1.000 & $0.371 \pm 0.014$ & $26.00 \pm 1.25$  & $5.89 \pm 2.01$ & $12300 \pm 1700$ & $1.2 \pm 0.3$ & $310 \pm 70$      \\
\multicolumn{1}{l|}{}                                &             &                  &                   & 0.833 & $0.378 \pm 0.024$ & $27.61 \pm 5.31$ & $5.77 \pm 3.43$ & $12100 \pm 2800$ & $1.1 \pm 0.4$ & $400 \pm 190$     \\
\multicolumn{1}{l|}{}                                & 16          & $1.800 \pm 0.05$ & $0.850 \pm 0.017$ & 1.800  & $0.349 \pm 0.013$ & $22.37 \pm 0.99$ & $4.89 \pm 1.30$ & $9800 \pm 1000$  & $1.2 \pm 0.2$ & $310 \pm 50$      \\
\multicolumn{1}{l|}{}                                &             &                  &                   & 1.250  & $0.363 \pm 0.017$ & $25.06 \pm 3.95$ & $5.24 \pm 2.13$ & $10700 \pm 1700$ & $1.1 \pm 0.3$ & $360 \pm 110$     \\
\multicolumn{1}{l|}{}                                &             &                  &                   & 1.000     & $0.369 \pm 0.015$ & $26.33 \pm 0.95$ & $6.31 \pm 2.37$ & $13400 \pm 2100$ & $1.4 \pm 0.3$ & $280 \pm 70$      \\ \hline
\multicolumn{1}{l|}{\multirow{11}{*}{\begin{turn}{90}Experiment II\end{turn}}} & 16          & $1.47 \pm 0.05$  & $0.625 \pm 0.017$ & 1.47  & $0.374 \pm 0.013$ & $21.71 \pm 1.31$ & $5.16 \pm 1.26$ & $10900 \pm 1100$ & $1.4 \pm 0.2$ & $280 \pm 50$      \\
\multicolumn{1}{l|}{}                                &             &                  &                   & 1.167 & $0.382 \pm 0.015$ & $22.88 \pm 1.06$ & $5.54 \pm 2.77$ & $12100 \pm 2100$ & $1.4 \pm 0.4$ & $300 \pm 110$     \\
\multicolumn{1}{l|}{}                                &             &                  &                   & 0.750  & $0.392 \pm 0.013$ & $25.93 \pm 2.18$ & $5.97 \pm 3.52$ & $13100 \pm 3200$ & $1.3 \pm 0.5$ & $350 \pm 160$     \\
\multicolumn{1}{l|}{}                                & 13          & $1.50 \pm 0.05$  & $0.70 \pm 0.017$  & 1.500   & $0.385 \pm 0.015$ & $21.56 \pm 1.80$ & $4.88 \pm 2.18$ & $10400 \pm 1900$ & $1.2 \pm 0.3$ & $340 \pm 110$     \\
\multicolumn{1}{l|}{}                                &             &                  &                   & 1.167 & $0.387 \pm 0.015$ & $23.98 \pm 1.10$ & $4.97 \pm 1.53$ & $10400 \pm 1200$ & $1.0 \pm 0.2$ & $380 \pm 80$      \\
\multicolumn{1}{l|}{}                                &             &                  &                   & 0.750  & $0.393 \pm 0.011$ & $28.18 \pm 1.49$ & $6.49 \pm 3.96$ & $14200 \pm 3700$ & $1.3 \pm 0.5$ & $350 \pm 170$     \\
\multicolumn{1}{l|}{}                                & 10          & $1.73 \pm 0.05$  & $0.917 \pm 0.008$ & 1.730  & $0.369 \pm 0.015$ & $23.58 \pm 4.32$ & $5.12 \pm 2.30$ & $10600 \pm 1900$ & $1.2 \pm 0.4$ & $350 \pm 120$     \\
\multicolumn{1}{l|}{}                                &             &                  &                   & 1.167 & $0.381 \pm 0.017$ & $26.46 \pm 1.23$ & $5.44 \pm 2.1$  & $11300 \pm 1700$ & $1.0 \pm 0.2$ & $390 \pm 100$     \\
\multicolumn{1}{l|}{}                                &             &                  &                   & 0.967 & $0.373 \pm 0.011$ & $27.69 \pm 2.62$ & $6.16 \pm 4.23$ & $13100 \pm 3700$ & $1.2 \pm 0.5$ & $370 \pm 220$     \\
\multicolumn{1}{l|}{}                                & 7           & $1.83 \pm 0.083$ & $1.167 \pm 0.008$ & 1.830  & $0.377 \pm 0.012$ & $25.76 \pm 1.77$ & $5.46 \pm 2.45$ & $11400 \pm 2000$ & $1.1 \pm 0.3$ & $370 \pm 110$     \\
\multicolumn{1}{l|}{}                                &             &                  &                   & 1.250  & $0.383 \pm 0.010$ & $28.17 \pm 1.40$ & $5.60 \pm 1.51$ & $11600 \pm 1200$ & $1.0 \pm 0.2$ & $410 \pm 70$\\
\hline

\end{tabular}
\end{table*}
As can be seen, decreasing the pressure at constant 
RF power resulted in an increase of both sound speeds and a slight increase 
of the interparticle distance. Assuming pure screened Coulomb interactions, it corresponded 
to an increase of the microparticle charges. 
Accordingly, the maximum frequency of the longitudinal dispersion relation (not shown here) increased 
when decreasing pressure making the out-of-plane mode and the in-plane mode 
closer to each other as the MCI pressure threshold, $p_{\rm MCI}$, was approached. 
No trend in the evolution of the coupling parameter 
$\kappa$ could be extracted due to the large error on the sound speeds, especially the 
transverse sound speed.

\section{Sheath profile above the powered electrode of a capacitively coupled radio-frequency discharges}\label{SheathModel}

Any modification of the discharge parameters ($p_{\rm Ar}$ and/or $P_{\rm W}$) 
has an impact on the plasma and therefore on the sheath properties.
It is then evident that since the microparticle monolayer levitates in the rf 
sheath above the powered electrode,  any changes of the discharge parameters also affect the 
monolayer properties. In this section, a simple model allowing 
calculation  of  the  sheath profile above the powered electrode in an asymmetric cc-rf 
discharge as a function of the discharge parameters is described. The model is then 
validated against available experimental data and then used to obtain the 
dependence of sheath parameters (sheath length, ion and electron densities, electric field) 
as  a function of the discharge parameters ($p_{\rm Ar}$ and $P_{\rm W}$).

\subsection{Description of the model}

In a cc-rf discharge, the rf frequency is generally much higher than 
the ion plasma frequency and much lower than the electron plasma frequency, 
i.e. $\omega_{\mathrm{pi}}\ll \omega_{\mathrm{rf}} \ll \omega_{\mathrm{pe}}$ 
where $\omega_{\mathrm{pi}},\ \omega_{\mathrm{pe}},\ \omega_{\mathrm{rf}}$ 
are the ion, electron and rf angular frequency, respectively. In that respect, 
the ions respond to the time-averaged sheath electric field. Moreover, in  
most of 2D complex plasma crystal experiments, the working pressure is around 1~Pa.
In this situation, ion collisions cannot be neglected  \cite{Lieberman1989}. The main 
assumptions of the following model are based on earlier work by M. Lieberman \cite{Lieberman1989}:
\begin{enumerate}
	\item{Collisional ion motion with a constant ion mean free path $\lambda_i$ in the sheath,}
	\item{Cold ions (ion temperature $T_\rmi\simeq0$),}
	\item{Inertia-less electrons that respond instantaneously to the electric field,}
	\item{No secondary electrons emitted from the electrode,}
	\item{No ionisation in the sheath.}
\end{enumerate}

In most experiments, asymmetric cc-rf discharges with grounded 
area much larger than the powered electrode area are used. 
Moreover, since the matching circuit very often contains a blocking capacitor, 
no direct-current  can flow in the external circuit.
Since the area of each electrode is different, the amount 
of collected ions and electrons during each rf cycle in also different 
 and results in the appearance of a self-bias voltage on 
 the powered electrode \cite{Raizer1995,Chandhok1998}.  Ideally, it would be necessary 
 to consider both sheath in
 front of both electrodes simultaneously \cite{Chandhok1998}.
 However, in order to simplify the problem, the following assumptions are made: 
 \begin{enumerate}
 
    \item{The voltage on the rf electrode is of the form:
            \begin{equation}
                V_\rmrf(t)=-V_{0}\cdot \cos(\omega_\rmrf t) -|V_{\rmdc}|
            \end{equation}
            The input values of $V_0$ and $V_\rmdc$ were discussed in Sec.~\ref{sec:Electrical}.
     }
     \item{At all time instants, the plasma must remain quasi neutral 
     and therefore the current going through the powered electrode sheath and the 
     anode sheath must be equal in intensity and of opposite signs. Following the method 
     described by Song et al.~\cite{Song1990}, 
     the time dependence of the plasma potential with respect to the grounded wall is approximated by:
     \begin{widetext}
     \begin{equation}
         V_\rmpl(t) =  T_\rme \ln{\Bigg[ \Bigg( 1+ \frac{A_\rmrf}{A_\rmg} 
         \exp{\Big( \frac{V_\rmrf (t)}{T_\rme} \Big)} \Bigg)\cdot 
         \Bigg(1+ \frac{A_\rmrf}{A_\rmg} \Bigg)^{-1} \Bigg]}
          + \frac{T_\rme}{2}\ln{\Bigg(\frac{m_\rmi}{2.3 m_\rme} \Bigg)}     
    \end{equation}
     \end{widetext}
     }
 \end{enumerate}
 
In the following, the plasma potential is taken as the reference potential ($V_\rmpl=0$). 
At any given instant of time, the potential of the rf electrode 
$V_\rmc$ with respect to the plasma  is:
\begin{equation}
    V_\rmc (t) =  V_\rmrf(t) - V_\rmpl(t).
\end{equation}
The time-dependant potential profile $V(z,t)$ in the 
sheath follows the Poisson equation:
\begin{align}
    \nabla^2 V(z,t) = & -\frac{e}{\epsilon_0}\Big(n_i(z) - n_e(z,t)\Big). \label{eq:poisson_time_dependant}
\end{align}
The boundary conditions are $V(0,t)=0$ and
$V(\ell_s,t=V_\rmc(t)$, where $\ell_s$ is the length of the ion sheath.
 
Since $\omega_{\mathrm{pi}}\ll \omega_{\mathrm{rf}}$, ions react only to the time-averaged electric 
field $\bar{E}=-\rmd \bar{V}/\rmd z$ where $\bar{V}$ is the time-averaged 
electric potential across the sheath of the powered electrode. Therefore, 
in the sheath, the ion particle conservation and momentum 
conservation equations are \cite{Godyak1990,Godyak1990a}:
 \begin{align}
    n_0 v_\rms & =  n_\rmi v_\rmi,\label{eq:ion1}\\
    M v_\rmi \frac{\rmd v_i}{\rmd z} + e \frac{\rmd \bar{V}}{\rmd z}+\frac{\pi M v_\rmi^2}{2 \lambda_\rmi}
   & =  0,\label{eq:ion2}
\end{align}
 where $n_0$ is the plasma density at the plasma sheath 
 boundary, $v_{\rmi}$ is the ion velocity at a given 
 position $z$ in the sheath, $v_\rms$ is the velocity of the 
 ions at the  ion sheath boundary ($z=0$), and $M$ is the ion mass.
 When taking into account collisions, $v_\rms$ is the modified Bohm velocity \cite{Godyak1990,Godyak1990a}:
 \begin{equation}
 	v_\rms=\Big( \frac{eT_\rme}{M}\Big)^{1/2} \cdot \Big(1+\frac{\pi \lambda_\rmde}{2\lambda_\rmi}\Big)^{-1/2},
 \end{equation}
 where $T_\rme$ is the electron temperature in eV, $e$ is the elementary charge, 
 $\lambda_\rmde=(\epsilon_0 T_\rme/n_0 e)^{1/2}$ is the electron Debye length.
 
\begin{figure}[htbp]
     \centering
     \includegraphics[scale=1.0]{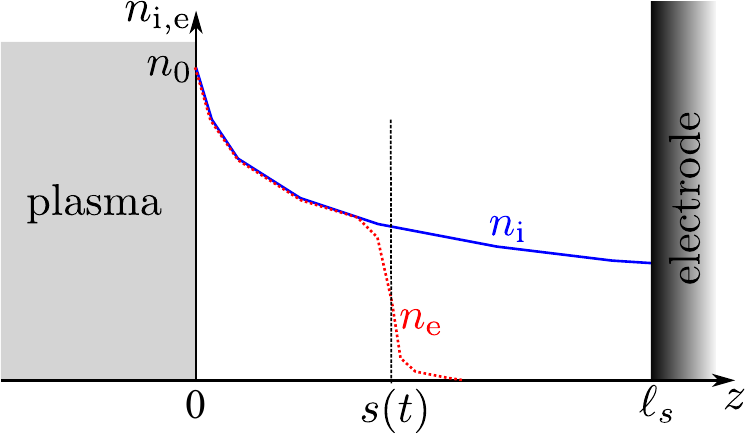}
     \caption{Schematic of a capacitive rf sheath at a time $0<t<\pi/\omega_\rmrf$. The dotted vertical line shows the position of the electron sheath boundary $s(t)$. For $z<s(t)$, $n_\rme\sim n_\rmi$ and for  $z>s(t)$, $n_\rme\sim 0$.}
     \label{fig:schematic_sheath}
 \end{figure}

 In order to calculate the sheath profile, one needs to make a few approximations. First of all, in order 
 to solve Eq.~\ref{eq:poisson_time_dependant}, one needs to know the exact position of the 
 time-varying sheath boundary. Indeed, depending on the voltage across the sheath, the electron sheath 
 boundary $s(t)$ will change and oscillate between the position of the ion sheath boundary to a minimum 
 distance to the electrode (see Fig.~\ref{fig:schematic_sheath}). 
 Generally, in cc-rf discharges, the amplitude of the rf voltage is much higher than the 
 electron temperature, i.e. $V_0\gg T_\rme$ and the electron Debye length is much smaller than 
 the sheath width $\ell_\rms$. It is thus clear that on the plasma side ($z<s(t)$), 
 the quasi-neutrality holds ($n_\rme\sim n_\rmi$), while on the sheath side ($z>s(t)$), 
 the electron density falls very rapidly ($n_\rme\sim 0$). Many studies dealing with the structure 
 of cc-rf sheath therefore use this approximation known as the step model which has been 
 extensively used in different models of cc-rf sheath 
 \cite{Lieberman1988,Lieberman1989,Godyak1990a,Chandhok1998,
 Sobolewski2000,Dewan2002,Gierling1998,Czarnetzki2013}. 
 In many of these models, the solution of the rf sheath relies on calculating the rf current through 
 the sheaths (powered and grounded) and assuming a zero net current 
 \cite{Lieberman1988,Lieberman1989,Godyak1990a,Chandhok1998,Dewan2002}. It allows one to calculate the 
 position of the electron sheath at all times and from there the other quantities such as the 
 average potential drop and average electric field in 
 the sheath can be derived. Calculations of these kind have 
 been performed for both symmetric \cite{Lieberman1988,Lieberman1989,Godyak1990a} 
 and asymmetric discharges \cite{Chandhok1998,Dewan2002}. It was nevertheless shown 
 that, in the case of symmetric 
 cc-rf discharges, these models were able to accurately 
 describe the current-voltage characteristics but showed 
 discrepancies in the potential and electric field profiles with respect to the exact 
 solution \cite{Gierling1998}. More recent models allow for very 
 accurate descriptions of the cc-rf sheath but 
 require input such as the maximum sheath extension, the ion mean free path and 
 the electron Debye length 
 \cite{Czarnetzki2013}. 
 The ion mean free path and the electron Debye length can be easily obtained from 
 discharge pressure, interferometry measurements and literature data. However, the maximum 
 sheath extension is not a parameter which is very often measured. Moreover, 
 the current and voltage waveforms  
 on the powered electrode are usually not monitored. However, in our experiments, 
 the self-bias voltage and the electron density were measured as a function of the 
 forward rf power and argon pressure over a wide range of 
 parameters which can be used to recover the 
 rf voltage amplitude $V_0$ using Eq.~\ref{eq:vdc}. The electron temperature can be 
 estimated using the uniform density discharge model \cite{Lieberman1994}:
 \begin{equation}
     \frac{K_{iz}(T_\rme)}{u_\mathrm{B}(T_\rme)}= \frac{1}{n_\rmg d_{eff}},
 \end{equation}
 where  $K_{iz}(T_\rme)=2.34\cdot10^{-14}\cdot T_\rme^{0.59}\cdot \exp(-17.44/T_\rme)$ is 
 the ionisation rate \cite{Lieberman1994}, $u_\mathrm{B}(T_\rme)=\sqrt{eT_\rme/M}$ is the Bohm velocity, 
 $n_\rmg$ is the neutral gas number density and $d_{eff}=0.5\cdot Rl/(Rh_l +lh_r)$ is the 
 effective plasma size for particle loss with $l$ the height of 
 the discharge chamber, $R$ its radius, $h_r=0.80\cdot(4+R/\lambda_{\rmi n})^{-1/2}$ and 
 $h_l=0.86\cdot(3+l/(2\lambda_{\rmi n}))^{-1/2}$ \cite{Lieberman1994} 
 and $\lambda_{\rmi n}$ the ion mean 
 free path in the plasma bulk. 
 
 The sheath profile can then be calculated using the following approximation 
 for the time-dependent Poisson equation:
 \begin{equation}\label{eq:poisson_Approx}
     \nabla^2 V(z,t) \simeq -\frac{e n_i(z)}{\epsilon_0} \Big( 1 - \exp{\frac{V(z,t)}{T_\rme}}\Big).
 \end{equation}
 As can be seen, this approximation lies between the ion 
 matrix sheath approximation ($n_i(z)=n_0$) and the 
 step model. It however allows us to solve the Poisson equation at anytime knowing the 
 voltage drop in the sheath. The mean electric field and 
 ion density profiles are then obtained in a recursive approach. 
Starting from the collision-less dc sheath profile with a potential drop equal to the maximum of 
$V_\rmc(t)$, a maximum sheath length and an initial ion density profile are calculated 
\cite{Lieberman1994,Moisan2006}:
\begin{eqnarray}
    \ell_s & = & \frac{\sqrt{2}}{3}\lambda_\rmde\cdot\Big( \frac{2\cdot\mathrm{max}(|V_c(t)|)}{T_\rme} \Big)^{3/4},\\
    \bar{V}(z) & = & - \mathrm{max}(|V_c(t)|)\cdot \Big(\frac{z}{\ell_s}\Big)^{4/3}, \\
    n_\rmi(z) & = & n_0\cdot\Big(1- \frac{2\bar{V}(z)}{T_\rme} \Big)^{-1/2}.
\end{eqnarray}
 Then the time-dependant potential profile over one rf period is 
 calculated by solving Eq.~(\ref{eq:poisson_Approx}) with \textsc{Matlab} and using 
 the following boundary conditions:
 \begin{align}
     V(0,t)  & =  0,\\
     V(\ell_s,t)  & = V_\rmc(t),
 \end{align}
 and new time-averaged potential and electric field are calculated. If the time-averaged electric 
 field on the plasma side ($z=0$) is lower than ${E_\mathrm{min}=(\pi/2)\cdot(T_\rme/\lambda_i)}$, 
 a value necessary to ensure monotonically decreasing ion density and increasing ion
 velocity in the sheath region, then the sheath boundary on the 
 plasma side is moved towards the wall until the average electric 
 field reaches $E_\mathrm{min}$ and 
 a new sheath length $\ell_s$ is obtained. Using Eqs.~(\ref{eq:ion1}) 
 and (\ref{eq:ion2}) with the boundary 
 conditions:
 \begin{align}
     n_\rmi(0) & =  n_0,\\
     v_\rmi(0) & = v_\rms,
 \end{align}
 a new ion density profile is obtained and the 
 time-dependant potential profile is calculated again. The procedure
 is repeated until it converges, which typically occurs in 10 to 15 iterations.

\subsection{Comparison to experimental data, validation of the model}

Schulze et al. have measured experimentally the time-dependant electric field of a cc-rf capacitive 
discharge in krypton at different pressures \cite{Schulze2008a}. Czarnetzki used this set of measurements 
to validate with success his advanced model of cc-rf sheath \cite{Czarnetzki2013}.

\begin{figure}[htbp]
    \centering
    \includegraphics[scale=1.0]{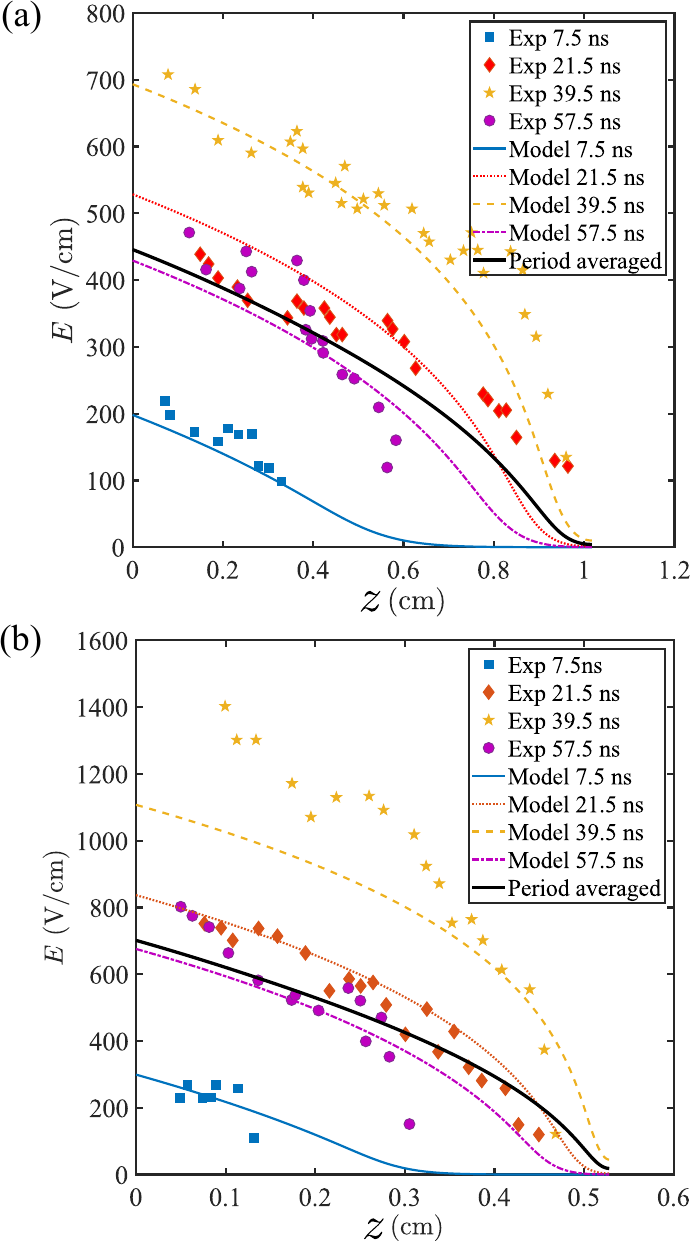}
    \caption{Experimental time-resolved electric field in krypton cc-rf discharge compared to 
    the calculated electric field using the model decribed in this article. The experimental data are 
    extracted from Ref.~\cite{Schulze2008a}. (a) $p_\mathrm{Kr}=1$~Pa  and 
    $P_\rmrf=8$~W, (b) $p_\mathrm{Kr}=10$~Pa  and $P_\rmrf=8$~W. The time instants 
    are given in the inset and in both plots the 
    plain black line corresponds to the time-averaged electric field.}
    \label{fig:comparison}
\end{figure}

Two sets of discharge parameters were used to validate our model:
\begin{enumerate}[label={(\alph*)}]
    \item{$p_\mathrm{Kr}=1$~Pa  and $P_\rmrf=8$~W, corresponding to $V_0\simeq250V$ and $V_\rmdc\simeq -250$~V (see Fig.~12 of Ref.~\cite{Schulze2008a}). As in Ref.~\cite{Czarnetzki2013}, the electron 
    temperature is fixed to $T_\rme=2.6$~eV and the plasma density at the sheath edge to 
    $n_0 = 2.0\cdot10^{15}$~m$^{-3}$. The grounded surface is assumed to be 100 times larger than the powered surface. The charge exchange cross section is $\sigma_{CX} = 40\cdot10^{-20}$~m$^2$ 
    \cite{Hause2013}.}
    \item{$p_\mathrm{Kr}=10$~Pa, and $P_\rmrf=8$~W, corresponding to $V_0\simeq225V$ and $V_\rmdc\simeq -225$~V (see Fig.~5 of Ref.~\cite{Schulze2008a}). As in Ref.~\cite{Czarnetzki2013}, the electron 
    temperature is fixed to $T_\rme=1.1$~eV and the plasma density at the sheath edge to 
    $n_0 = 7.3\cdot10^{15}$~m$^{-3}$. The grounded surface is assumed to be 100 times larger than the powered surface.  The charge exchange cross section is $\sigma_{CX} = 40\cdot10^{-20}$~m$^2$
    \cite{Hause2013}.}
\end{enumerate}

In Fig.~\ref{fig:comparison}, the model results are compared to the experimental data of 
Schulze et al. \cite{Schulze2008a}. As can be seen, the agreement is not perfect and the model 
tends to slightly underestimate the electric field, especially at large pressure. 
However, one should note that 
the length of the sheath is recovered within a few percent and that the electric field 
values are in respectable agreement with the measured ones.

We can therefore conclude that our simple model is able to catch the main features of the collisional 
cc-rf sheath and can reasonably estimate the main sheath parameters as a function of position for 
an adequate range of discharge parameters.

\subsection{Calculated sheath profile}\label{CalculatedProfiles}

\begin{figure*}[htbp]
    \centering
    \includegraphics[scale=0.67]{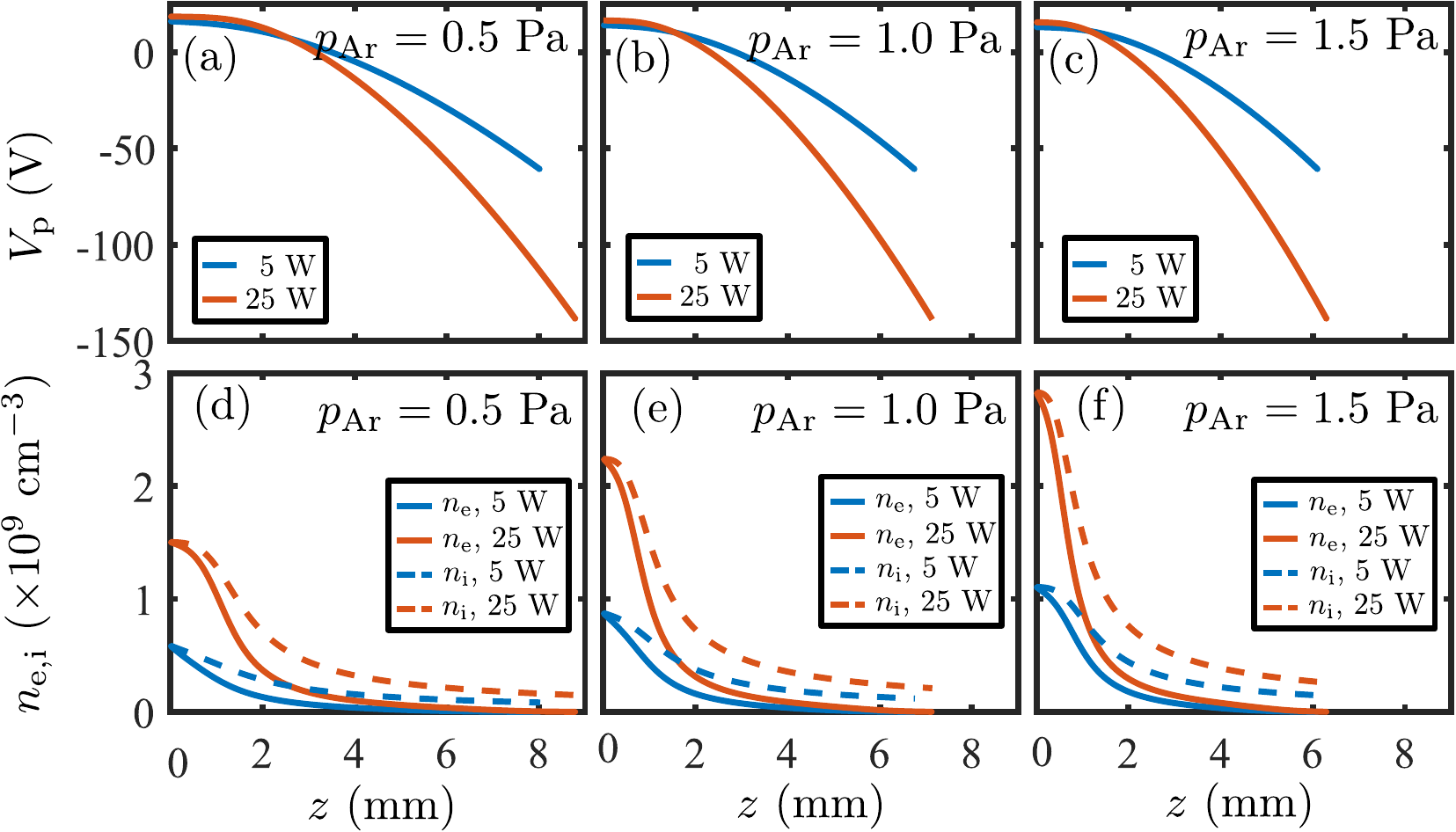}
    \caption{(Colour online) (a-c) Potential and (d-f) density profiles 
    for different pressures and powers. Note that potentials are plotted with 
    respect to ground (i.e. $V_{\rm p}(0) \neq 0$). }
    \label{fig:simulated_sheath}
\end{figure*}

The model was then used for conditions relevant to our experimental conditions.
The argon pressure was varied between 0.5~Pa and 2~Pa by steps of 0.05~Pa and the rf 
power was varied between 5~W and 25~W by steps of 1~W. 
In Fig.~\ref{fig:simulated_sheath} the sheath potential 
(Figs.~\ref{fig:simulated_sheath}(a-c)) 
and the electron and ion 
densities (Figs.~\ref{fig:simulated_sheath}(d-f)) are shown as a function of position  for 
different pressures and rf powers.

As can be seen in the top row of Fig.~\ref{fig:simulated_sheath}, increasing pressure 
at constant rf power results in shorter sheath length for a similar potential drop (the sheath 
length is $\sim8-9$~mm at $p=0.5$~Pa and drops down to $\sim 6-6.5$~mm at $p=1.5$~Pa). Therefore,   
the vertical electric field is weaker at lower pressures. One can therefore expect a weaker 
vertical confinement for microparticles levitating in the rf sheath at low pressure for a given rf power.
Increasing the rf power at constant pressure results in a slight increase of the sheath 
length but larger plasma potential drops due to increasing self-bias voltage 
(the potential drop is $\sim50$~V at $P_W=5$~W and 
 $\sim 140$~V at $P_W=25$~W). Therefore, a stronger vertical electric field and 
 a stiffer vertical confinement for microparticles levitating in the rf sheath at 
high rf power are expected.

 Both $P_{\rm W}$ and $p_{\rm Ar}$ have an impact on plasma density. 
 We know that at a given pressure, the plasma density 
is increasing almost linearly with rf power (see Fig.~\ref{fig:ne_exp}). Since the sheath 
length is weakly dependant on rf power (see previous paragraph), the gradients of ion and electron 
densities in the sheath increase with rf power as can be seen in the bottom row of Fig.~\ref{fig:simulated_sheath}. At high pressure, the plasma density 
is higher (due to collisions), the sheath length is short resulting in steep density gradients. 

Using the calculated sheath profiles, the dependence of dust particle charge number, $Z_\rmd$, 
levitation height, $z_{\rm lev}$ and vertical resonance frequency, $f_{\rm v}$ on discharge 
parameters is explored in the next section.

\section{Microparticle properties and mode-coupling instability}\label{MCI}

In this section, the influence of discharge parameters on $Z_\rmd$, 
$z_{\rm lev}$, and $f_{\rm v}$ is investigated. Then, 
the threshold of MCI in 2D complex plasma crystal as a function of discharge 
parameters is studied.

\subsection{Dust particle charge, levitation height and vertical resonance frequency}\label{ParticleCharges}

With the calculated sheath profile as a function of rf power and pressure, the 
equilibrium dust particle charge and levitation height can be calculated. At any point in the 
sheath, the ion and electron densities are known allowing the calculation of 
ion and electron currents, $I_\rmi$ and $I_\rme$ respectively onto the 
microparticle surface in a collisional plasma \cite{Zobnin2008,Khrapak2009}:
\begin{widetext}
\begin{align}
    I_\rme  = & (-e)\cdot \sqrt{8\pi}r_\rmd^2 n_\rme v_{T_\rme} \exp(-\tilde{\varphi}), \\
    I_\rmi  = & (e)\cdot\sqrt{8\pi} r_\rmd^2 n_\rme v_{T_\rmi} 
    \Big(1+\tilde{\varphi} \Big) \times \nonumber \\ 
     & \left[ 1+  \frac{\tilde{\varphi}\left(\frac{T_\rme}{T_\rmi}\frac{r_\rmd}{\lambda_\rmi}\right)}{0.07
           +2\left(\frac{r_\rmd}{\lambda}\right) + 2.5 \left(\frac{r_\rmd}{\lambda_\rmi}\right)+
           \left[ 0.27 \Big(\frac{r_\rmd}{\lambda} \Big)^{1.5} 
           + 0.8 \Big( \frac{r_\rmd^2}{\lambda_\rmi \lambda} \Big)\right] \frac{T_\rme}{T_\rmi} \tilde{\varphi} 
           +\frac{0.4\left( \frac{r_\rmd}{\lambda_\rmi}\right)^2\left(\frac{T_\rme}{T_\rmi}\right)\tilde{\varphi}}
           {1-0.4\left(\frac{r_\rmd}{\lambda_\rmi}\right)}
           } \right],
\end{align}
\end{widetext}
where $r_\rmd$ is the radius of the microparticle, $\tilde{\varphi}$ is 
the normalised microparticle floating potential 
and $\lambda=\lambda_{\mathrm{D_i}}/\sqrt{1+(\lambda_\mathrm{D_i}/\lambda_\mathrm{D_e})^2}$ 
is the linearised Debye length at the considered position in the sheath 
with $\lambda_{\rm D_{i,e}}=(\epsilon_0 T_{\rm i,e}/n_{\rm i,e} e)^{1/2}$. 
At equilibrium $I_\rmi + I_\rme=0$ and the normalised  
floating potential $\tilde{\varphi}$ can be obtained for any position in the sheath (ions 
are assumed to be in thermal equilibrium with the background gas, $T_{\rm i}=0.03$~eV). 
The microparticle  charge number $Z_\rmd=Q_\rmd/e$ where $Q_d$ 
is the microparticle charge is then obtained:
\begin{equation}
    Z_\rmd= -4\pi\epsilon_0 r_\rmd T_\rme \tilde{\varphi}/e.
\end{equation}
Since $r_\rmd$ and therefore the mass $m_\rmd$ of the microparticle are known 
and since the sheath electric field profile has been calculated, 
the equilibrium levitation height $z_\mathrm{lev}$ and the vertical resonance 
(confinement) frequency $f_\mathrm{v}=\omega_v/(2\pi)$ can be obtained using 
the following equations:
\begin{align}
    Q_\rmd E(z_\mathrm{lev})=-m_\rmd g, \\
    \omega_v=\sqrt{\frac{|Q_\rmd |}{m_\rmd} \frac{\partial E(z)}{\partial z}}_{|z=z_\mathrm{lev}}.
\end{align}

\begin{figure*}[htbp]
    \centering
    \includegraphics[width=0.99\textwidth]{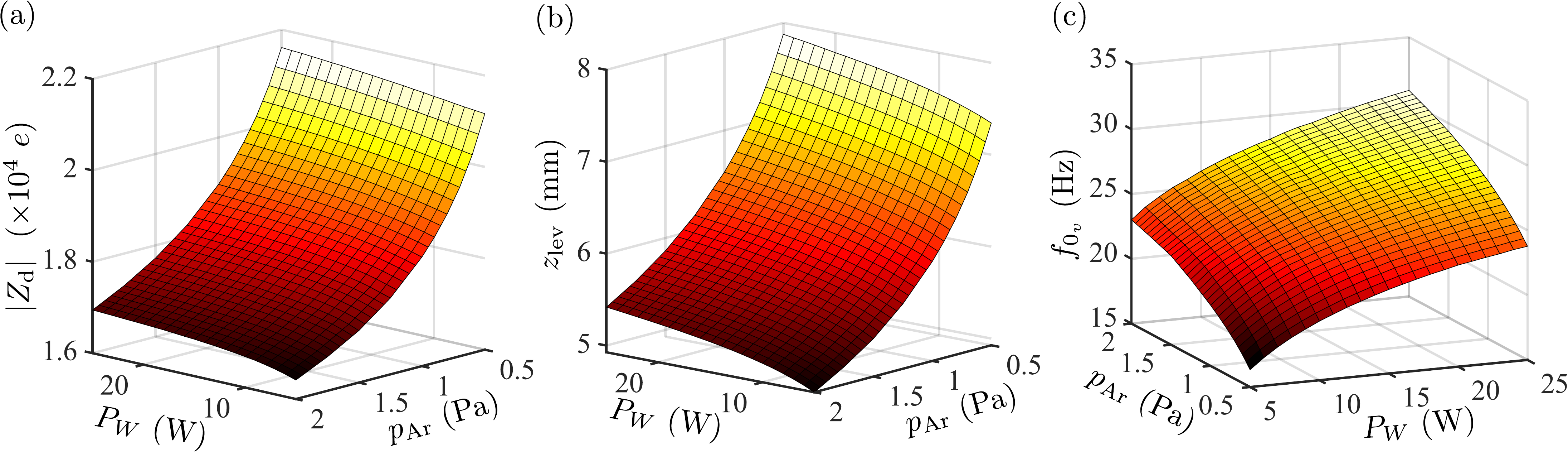}
    \caption{(a) Equilibrium particle charge. (b) Equilibrium levitation height. (c) 
    Vertical resonance frequency. The microparticle has a diameter $d=9.16\ {\rm \mu m}$ 
    and its mass density is $\rho_m=1510\ {\rm kg \cdot m^{-3}}$ (MF density).}
    \label{fig:sheath_particle_resuls}
\end{figure*}

In Fig.~\ref{fig:sheath_particle_resuls}, the results of 
the calculation of  levitation height, particle charge at equilibrium 
position and vertical resonance frequency as a function of rf power and background 
argon pressure are shown.
The obtained equilibrium particle charges are between $-2.2\cdot 10^4e$ and $-1.6\cdot 10^4e$ (see 
Fig.~\ref{fig:sheath_particle_resuls}(a)), 
which is slightly larger than the measured one (see Table.~\ref{tab:crystal_parameters}) but are of the 
same order of magnitude.
One can see that at a given argon background pressure, $Z_\rmd$ and $z_{\rm lev}$ 
are only slightly increasing with rf power. Moreover, 
the microparticles levitate quite high in the sheath near the sheath edge (for example at 1~Pa, 
the sheath length is around $\sim7$~mm [see Fig.~\ref{fig:simulated_sheath}] and 
the levitation height is around $\sim 6.5$~mm [see Fig.~\ref{fig:sheath_particle_resuls}(b)]). 
This is due to the fact that even though the ion and electron densities are increasing 
with rf power, the electron temperature is mostly set by the background gas pressure. Since $T_\rme$ 
has a greater influence than $n_\rme$ on the magnitude of the electron current onto 
a microparticle and therefore the microparticle charge, the microparticle 
equilibrium charge depends only weekly on rf power. Since at equilibrium levitation height, 
the gravity force is compensated 
by the electric force and, as seen previously, the sheath length being only 
weakly dependant on rf power, the particle levitation height will 
only slightly change. 

The vertical resonance frequency is increasing significantly with both rf power and 
argon background pressure (see Fig.~\ref{fig:sheath_particle_resuls}(c)). In both cases, this is
due to steeper electric field when increasing rf power (resulting in larger self-bias) and 
when increasing the background argon pressure (resulting 
in shorter sheath). Therefore, decreasing pressure 
or rf power results in weaker vertical confinement and favours MCI for 2D complex plasma 
crystals. On the contrary, high pressure and high rf power favour crystallisation of complex plasma 
monolayer by providing stiffer vertical confinement. More details are given in the next subsection. 

\subsection{Mode-coupling instability}

It is well known that MCI can occur in 2D complex plasma crystals 
\cite{Zhdanov2009, Couedel2010, Couedel2011} and in fluid complex plasma 
monolayers \cite{Ivlev2014,Yurchenko2017,Couedel2018}. In Sec.~\ref{expMCI}, it was 
demonstrated experimentally that decreasing
pressure and/or rf power leads to MCI in 2D complex plasma crystals 
and that the MCI threshold and crystallisation threshold have defined trends. 

One of the simplest ways to simulate MCI is to use the point-charge wake model 
\cite{Zhdanov2009,Couedel2011,Rocker2012}. In this model, the ion wake is approximated 
by a positive point charge $q_w$ downstream of each microparticle at a distance $\delta_w$ below the 
microparticle. Then each microparticle interacts with the other microparticles and their 
respective ion wakes through screened-Coulomb (Yukawa) interactions. The interparticle 
distance in the 2D crystal is set at a fixed value $\Delta$. In the calculation, 
the screening length is taken as the electron Debye length, $\lambda_{\rm D_e}$ 
at the equilibrium leviation height.
To compute the 
different lattice modes in the 2D complex plasma crystals, one can calculate the dynamical matrix
which has the form \cite{Zhdanov2009,Couedel2011,Rocker2012}:
\begin{equation}
    \mathbf{D}=\begin{pmatrix}
    \alpha_h - \beta & 2\gamma & \imath \sigma_x\\
    2\gamma &\alpha_h + \beta & \imath \sigma_y\\
    \imath \sigma_x & \imath \sigma_y & \omega_v^2 - 2 \alpha_v\\
    \end{pmatrix},
\end{equation}
where $\alpha_{h,v}$, $\beta$ and $\gamma$ are the dispersion elements,  
$\sigma_{x,y}$ are the coupling terms between the in-plane and out-of-plane modes
due to the particle-wake interaction. Detailed expressions of these different terms
as functions of $Q_d$, $\lambda_{\rm D_e}$, $q_w$, $\delta_w$, $\Delta$  
and $\mathbf{k}$, where $\mathbf{k}$ is the wave vector of the considered lattice mode can be 
found in Refs.~\cite{Zhdanov2009,Couedel2011,Rocker2012}. The dispersion relation of 
the microparticle lattice modes are obtained for any $\mathbf{k}$ by solving
$\mathrm{det}[\mathbf{D} - \omega(\omega +\imath \nu)\mathbf{I}]=0$,  
where $\mathbf{I}$ is the unit matrix, so that $\omega(\omega +\imath \nu)$ are 
the eigenvalues of the dynamical matrix, where $\nu$ is the  
damping rate due to neutral gas friction.

\begin{figure}[htbp]
    \centering
    \includegraphics[scale=1.0]{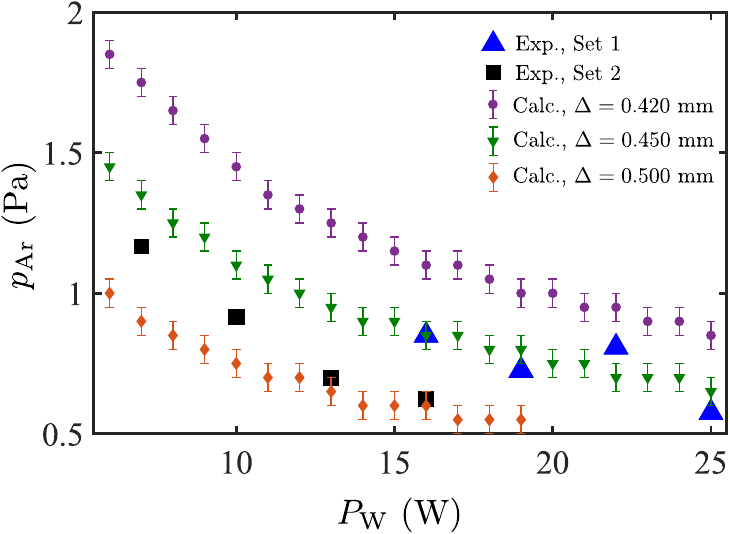}
    \caption{(Colour online) Pressure at which the crossing of the in-plane longitudinal mode and the out-of-plane mode occurs as a function of rf power for different interparticle 
    distances $\Delta$ and without ion wakes. The experimental MCI 
    pressure thresholds are shown for comparison.}
    \label{fig:calc_nowake}
\end{figure}

It is known that the mode coupling instability in a 2D complex plasma crystal is triggered 
when the out-of-plane mode and the in-plane longitudinal mode 
cross \cite{Zhdanov2009,Couedel2011} except in very specific experimental 
conditions where the ion wake charges are really strong allowing 
coupling of the modes without direct crossing \cite{Meyer2017}. 
Firstly, the influence of the interparticle distance on mode 
crossing is investigated in the case 
where ion wakes are ignored ($\sigma_{x,y}=0$). Using the results of Sec.~\ref{CalculatedProfiles} and 
Sec.~\ref{ParticleCharges},  the minimum 
pressure for out-of-plane and in-plane longitudinal mode crossing as a function of power were 
calculated for different $\Delta$ (420~${\rm \mu m}$, 450~${\rm \mu m}$ and 500~${\rm \mu m}$). 
The results are presented in Fig.~\ref{fig:calc_nowake}. As can be 
seen, the  out-of-plane and in-plane longitudinal mode crossing minimum 
pressure decreases when increasing the rf power. This can be 
understood due to the competing effects of 
the pressure decrease which makes sheath longer and therefore 
weaken the confining electric field and 
the rf power increase that increases the self-bias of the powered 
electrode and makes the vertical 
confinement stiffer. Moreover, at low pressure the microparticle 
charge is larger due to a higher 
electron temperature, which for a fixed 
interparticle distance renders the interaction force greater 
and therefore increases the maximum frequency 
of the longitudinal in-plane mode and makes mode 
crossing easier. In addition, one can see that the pressure 
threshold is systematically lower for larger $\Delta$. This is due to the weaker 
interparticle interaction 
which reduces the maximum frequency of the longitudinal 
in-plane mode and the minimum frequency 
of the out-of-plane mode. Finally, we can see that 
experimental points follow roughly the same trend. This 
shows that the sheath profile (which depends on $p_{\rm Ar}$ and $P_{\rm W}$) 
plays a major role 
in the crossing of the modes and the triggering of MCI in 2D complex plasma crystals.

\begin{figure}[htbp]
    \centering
    \includegraphics[scale=1.0]{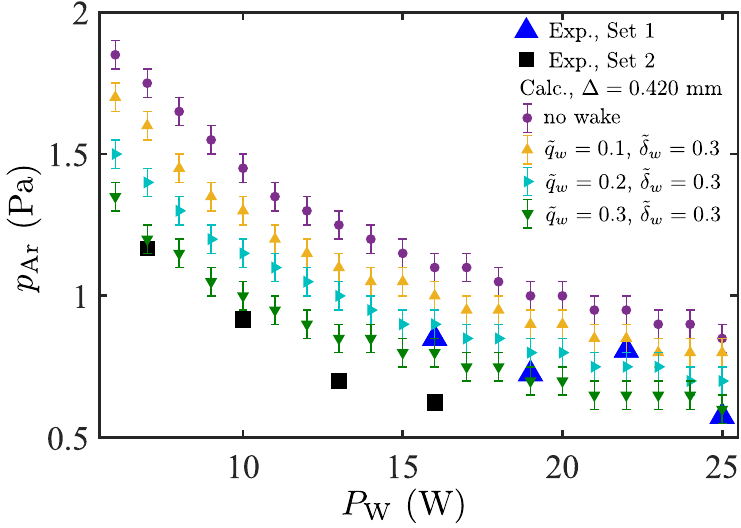}
    \caption{(Colour online) MCI pressure threshold as a function of rf power for an interparticle 
    distance $\Delta=420\ {\rm \mu m}$ and different ion wake charges in the framework 
    of the point-charge wake model \cite{Zhdanov2009}. The experimental MCI 
    pressure thresholds are shown for comparison.}
    \label{fig:calc_wake}
\end{figure}

In a second step, the interparticle distance was fixed to 
$\Delta=420$~${\rm \mu m}$ and point-charge 
wakes were considered. Only the value of the wake charge 
was varied ($0.1\leq \tilde{q}_w=q_w/Q_d \leq 0.3$),  
while the distance to the microparticle is fixed ($\tilde{\delta}_w=\delta_w/\lambda=0.3$). Then, 
the pressure threshold for the crossing of the modes (and 
the triggering of MCI) were calculated again.
The results are presented in Fig.~\ref{fig:calc_wake}. 
As can be seen, the main trends are still there:
when increasing the rf power, the threshold pressure 
decreases. However, the ion wakes have a significant 
effect. Increasing the ion wake charge significantly decreases 
the pressure threshold at a given rf power.
The wake charges are unknown in the experiment and 
their overall effect on the measured pressure 
threshold are difficult to estimate. However, one can see in 
Figs.~\ref{fig:calc_nowake} and \ref{fig:calc_wake} that the experimental 
threshold values are not exactly aligned with 
the calculated one. Note again that the experimental particle charges are slightly 
smaller than the calculated ones and $\Delta$ used 
in the calculation is slightly larger than the 
experimental one (which is in addition not a constant for a 
given $P_{\rm W}$ but slightly decreases with 
$p_{\rm Ar}$. In order to calculate the true MCI pressure threshold, 
it would be necessary to take into account the 
evolution of the horizontal confinement as a function of  $p_{\rm Ar}$ and $P_{\rm W}$ 
as well as self-consistent wake parameters for the calculation of the 
 lattice modes. A more sophisticated model 
for the ion wake such as those described in Refs.~\cite{Rocker2012,Rocker2012a,Kompaneets2016} 
is also needed as the wake model has a strong influence 
on the shape of the lattice modes \cite{Rocker2012}.

To conclude, let us discuss the trend for the crystallisation pressure observed in 
Fig.~\ref{fig:exp_stability}. As it was shown in 
Sec.~\ref{expMCI}, the crystallisation pressure 
follows the same trend as the MCI threshold pressure: it decreases with increasing power.
This can be roughly explained through the modification of the sheath profile. When 
the $P_{\rm W}$ increases, the 
sheath electric field and therefore the vertical confinement become stronger. However, 
when in the fluid phase, the crossing of the out-of-plane mode with the longitudinal 
in-plane mode always occurs and MCI will exist unless damping due to neutral friction is high 
enough to suppress it \cite{Ivlev2014}. At constant rf power, 
the microparticle charges decrease 
when increasing pressure (see Fig.~\ref{fig:sheath_particle_resuls} 
and Tab.~\ref{tab:crystal_parameters}). 
Therefore, the longitudinal in-plane mode is not as 
steep (the longitudinal sound speed decreases, 
see Tab.~\ref{tab:crystal_parameters}) and it will result in smaller instability 
growth rate. It was also shown in  Ref.~\cite{Ivlev2014}, in the 
framework of the wake-layer model, that when the vertical confinement 
frequency is increased above the threshold for mode crossing 
for MCI in a 2D crystal with the same 
microparticle parameters as in the fluid layer (same microparticle charges and same 
microparticle number density), the instability 
growth rate decreases as $\sim \exp(-\frac{1}{2}\omega_v^2 \delta)$. 
The crystallisation threshold 
is therefore also influenced by the wake parameters. Since these parameters are unknown, the 
crystallisation pressures were not calculated but ion wakes are expected to have a 
non-negligible effect as for the 
MCI threshold pressure in 2D complex plasma crystals. 
In future studies, the use of a self-consistent 
wake model such as in Ref.~\cite{Kompaneets2016} in 
addition to proper modelling of the horizontal 
confinement as a function of discharge parameters should enable us to derive self-consistently 
the MCI threshold and crystallisation pressures.

\section{Summary and Conclusion}\label{Conclusion}

In this article, the MCI threshold pressure and crystallisation pressure of monolayer 
complex plasma levitating in the sheath of  
an argon cc-rf discharge were investigated. It was found that the stability of a crystalline 
microparticle monolayer increases with pressure and with the rf power:  
at constant rf power, the higher the rf power the lower the crystallisation and MCI 
threshold pressures; 
at constant background argon pressure, the higher the pressure the lower rf power 
for crystallisation of the monolayer and for the triggering of MCI. 
It was also found by measuring the longitudinal and transverse sound speeds in 
two dimensional complex plasma crystal that at constant rf power, 
lower pressure resulted in larger sound speeds and therefore larger microparticle charges. These 
results are in agreement with previous observations of the MCI \cite{Couedel2010,Couedel2011}.

A simple rf sheath model was developed to calculate the evolution of the rf sheath profile 
as a function of the input rf power and the background argon pressure. Using experimental measurements
of the electron density in the plasma bulk and the powered electrode self-bias as input parameters, 
the model was able to show that the argon pressure and the rf power have significant influence on 
sheath profiles. Using the calculated sheath parameters, 
the microparticle equilibrium charge, the equilibrium levitation height  and 
the vertical confinement frequency were also calculated. It was found that: (i) a power increase at 
constant pressure leads to increase of $Z_{\rm d}$, $z_{\rm lev}$ and $f_{\rm 0_v}$ and (ii) 
a pressure increase at constant rf power leads to decrease of $Z_{\rm d}$, $z_{\rm lev}$ and 
an increase of $f_{\rm 0_v}$. Combined with the point-charge wake model \cite{Zhdanov2009}, 
the trends for MCI threshold pressure could be reproduced 
(i.e. decreasing the pressure at constant power would lead to the 
crossing of the longitudinal in-plane mode and the out-of-plane mode 
and trigger MCI, lower rf power results in higher MCI threshold pressure). 
However, it was also shown that 
the MCI threshold is  sensitive to interparticle distance and  wake 
parameters.

Compared to earlier studies, a better integration of the experimental results and theory 
was achieved by operating with actual experimental control parameters such as the 
gas pressure and the discharge power. However, future studies will 
concentrate on the use of a self-consistent wake model \cite{Kompaneets2016} to 
study  the influence of ion wakes on dust equilibrium position and vertical resonance frequency.
The influence of the plasma parameters on horizontal confinement should also be taken into account in 
order to model self-consistently the interparticle distance in monolayer 
complex plasma enabling more detailed investigations of 
fluid MCI and the crystallisation threshold pressure.

\begin{acknowledgments}
The authors would like to thank S.~Zhdanov for fruitful comments and discussions. 
The authors would also like to thank H.~Thomas for his careful reading of our manuscript.
L. Cou\"edel acknowledges the support of the Natural Sciences and 
Engineering Research Council of Canada (NSERC), RGPIN-2019-04333. 
\end{acknowledgments}

%

\end{document}